\documentclass{aa}
\usepackage{natbib}
\usepackage{graphicx}
\usepackage{txfonts}
\begin{document}

\title{An active asteroid belt causing the UX Ori phenomenon in RZ\,Psc}
\author{W.J. de Wit\inst{1} \and V.P.\,Grinin\inst{2,3} \and I.S.\,Potravnov\inst{2,3} \and D.N.\,Shakhovskoi\inst{4} \and A. M\"{u}ller\inst{1} \and M. Moerchen\inst{1,5,6}}
\institute{European Southern Observatory, Casilla 19001, Santiago 19, Chile\and
Pulkovo Astronomical Observatory, Russian Academy of Sciences, 196140, Pulkovo, St.\,Petersburg, Russia \and
The Sobolev Astronomical Institute, St. Petersburg University, Petrodvorets, St. Petersburg, Russia \and
Crimean Astrophysical Observatory, P. Nauchny, Crimea, Ukraine\and
Space Telescope Science Institute, 3700 San Martin Drive, Baltimore, MD 21218, USA\and
Leiden Observatory, PO Box 9513, 2300 RA Leiden, The Netherlands}

\offprints{W.J. de Wit, \email{wdewit@eso.org}}

\date{Received\,/\,Accepted}

\titlerunning{The dust of RZ\,Psc}

\authorrunning{de Wit et al.}

\abstract{We report the discovery of mid-IR excess emission in the
  young object RZ\,Psc. The excess constitutes $\sim$8\% of its
  L$_{\rm bol}$, and is well fit by a single 500\,K blackbody implying a dust
  free region within $\sim0.7$\,AU,  for optically thick dust. The object
  displays dust obscuration events (UXOR behaviour) with a timescale that suggests 
  dusty material on orbits of $\sim0.5$\,AU.  We also report a 12.4
  year cyclical photometric variability which can be interpreted as due to
  perturbations in the dust distribution. The system is characterized by a high inclination, marginal extinction (during
  bright photometric states), a single temperature for the warm dust, and an age estimate which puts the star beyond the formation stage. 
  We propose that the dust occultation events present a dynamical view of an
  active asteroid belt whose collisional products sporadically obscure the
  central star.}

\maketitle

\section{Introduction}
In the course of the formation of a star, an equatorial disk
is present whose purpose evolves from an angular momentum
re-distributor facilitating star growth to a planet builder.  Dust
grains in the disk change chemically \citep{2001A&A...375..950B} and physically through
growth \citep{2004Natur.432..479V} and
collisions \citep{2007ApJ...663..365W}. Observations demonstrate on the
one hand the presence of dust disks by their tell-tale thermal
infrared spectrum, whereas on the other hand hundreds of mature planetary
systems are now known. Yet, how the dust disk evolves to a planetary system is not well
understood \citep{2011ARA&A..49...67W}.

RZ Psc is a solar-type star (K0\,IV, Herbig 1960\nocite{1960ApJ...131..632H}) and well-known 
for its brightness variability with time. The
variability has all the hallmarks of the so-called UXOR variability seen among pre-main
sequence stars. They sporadically have photometric minima with
amplitudes of $\Delta V \approx 2^m-3^m$ which last from a few days up
to a few weeks. During a minimum, the UXOR displays bluer optical colours
and an increased linear polarization due to an increased contribution
by scattered light off small dust grains (e.g. Grinin 1988\nocite{1988SvAL...14...27G}; Grinin et al. 1991\nocite{1991Ap&SS.186..283G}). This type of
variability is strictly associated with the occultation of the star by dust in 
the optically thick accretion disks of stars younger than 10 million years
(Dullemond et al. 2003\nocite{2003ApJ...594L..47D}). The age of RZ\,Psc is estimated to
be approximately a factor three older (Grinin et al. 2010, hereafter paper\,I)\nocite{2010A&A...524A...8G}.
Correspondingly, the star does not display the benchmark properties of young
stars like ionized gas transitions or excess emission by hot (1500\,K)
dust (Bertout 1989; Paper\,I)\nocite{1989ARA&A..27..351B,2010A&A...524A...8G}. Rz\,Psc is therefore 
enigmatic because of a variability normally caused by optically thick accretion disks.
In this follow-up study, we report that the object has one of the strongest 
infrared excesses observed to date and provide a detailed periodicity analysis 
of the optical variability.

\section{The observational data}
In order to shed light on the cause for the star's continuum emission
variability, peculiar for its age, we explored the mid-infrared and
far-infrared wavelength region by means of data obtained by a variety
of infrared satellite missions and ground-based surveys, {\it viz.}
{\it Wide-field Infrared Survey Explorer} (WISE, Wright er al. 
2010\nocite{2010AJ....140.1868W}), {\it Infrared Astronomical Satellite}
(IRAS), {\it AKARI} \citep{2010A&A...514A...1I} and the {\it Two
  Micron All Sky Survey} (2MASS, Cutrie et al. 2003\nocite{2003yCat.2246....0C}). 
We also present new $V$-band photometry taken at the Crimean Astrophysical
Observatory (CrAO) extending the star's light curve to March 2012
(see Fig.\,2).

Representative $UBVRI$ magnitudes for the star's photosphere are taken from 
the mean value of the three brightest measurements, assuming that these
are least affected by circumstellar extinction.  The 2MASS near-IR magnitudes in
the $JHK$ bands reveal no change with respect to the measurements by
Glass \& Penston (1974\nocite{1974MNRAS.167..237G}), $H=9.7\pm0.2$ and
$K=9.78\pm0.1$. The WISE point source coincides with RZ\,Psc
within 0.3\arcsec\,and is the only catalogued WISE source within a 23\arcsec\,radius
of our target. We also note that the IRAS 12\,$\mu$m flux is
consistent with the WISE flux measurement at 11.6\,$\mu$m.  Finally, the IRAS
Faint Source Reject Catalog contains flux upper limits at 60\,$\mu$m and
100\,$\mu$m.

\section{Results and discussion}
In paper\,I, we estimated an age for RZ\,Psc of 30$-$40\,Myr
based on its kinematics and Lithium absorption. An age constraint
using the star's space motion was performed, exploiting the high
galactic latitude of $b\sim35\degr$, and assuming that the star
formed at $b=0\degr$. Clearly this is an uncertain exercise, but
the result is consistent with the observed Li equivalent width, which
is consistent with an age between 10 and 70\,Myr.
Recently, we measured a $v_{\rm rad}$ of $\rm -2.0 \pm
1.5\,km\,s^{-1}$ (Potravnov \& Grinin 2013\nocite{}), consistent with the
ones reported in Shevchenko et al. (1993), but in disagreement
with the literature value ($\rm -11.5\,km\,s^{-1}$ ) used in Paper\,I.
Using the new value, the kinematic age changes slightly to
$t_{\rm kin} \sim 25 \pm 5$\,Myr (standard error). Arguments in support of a system
older than 10\,Myr include the absence of interstellar dust
extinction towards RZ\,Psc (the star lies far from active star formation regions) and the lack of $H\alpha$ emission. In the
following, we will discuss the star's properties assuming that the star has finished its formation
process, because the age constraints available exceed the
characteristic time for optically thick accretion disks found
for classical T\,Tauri stars, but keeping in mind that the system
is nonetheless relatively young.

\subsection{The Spectral Energy Distribution}
The resulting spectral energy distribution (SED) is presented in
Fig.\,1.  It reveals the star's photospheric emission at visual
wavelengths and strong excess emission dominating the total radiation
from 3\,$\mu$m onwards.  We note that the photospheric fit does not
require an extinction correction (Kaminskii et
al. 2000\nocite{2000ARep...44..611K}). IR excess emission is
observed among a small fraction of main-sequence stars, {\it viz.} the
debris disk objects \citep{2008ARA&A..46..339W}. The emission is caused by
warm dust particles released in collisions between planetesimals in belts similar to the asteroid and Kuiper belt
of our Solar System \citep{2008ARA&A..46..339W}.  However, the excess
emission observed in RZ\,Psc is found to have two outstanding
properties that sets it apart from regular debris disk systems.

\begin{figure}
\includegraphics[width=6.cm, height=8cm,angle =90]{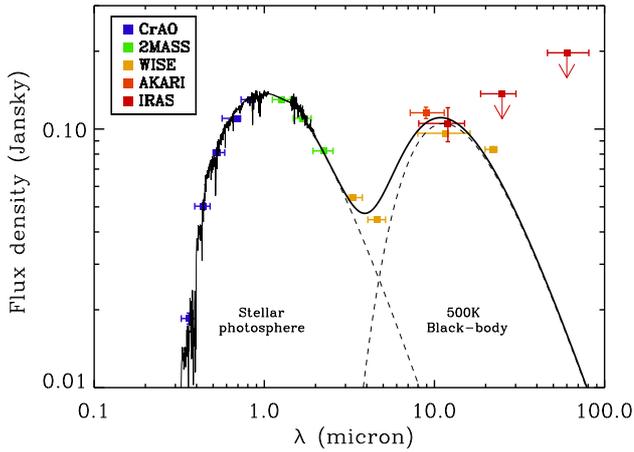}
\caption{\label{1} The SED of RZ\,Psc.
  A K0\,IV model photosphere \citep{1998PASP..110..863P} fits the optical and
  near-IR measurements shortward of 3\,$\mu$m. Longward of 3\,$\mu$m,
  the excess emission is markedly dominant and contributes
  8\% to the total source luminosity. The solid black line
  corresponds to the total flux of stellar photosphere plus a single
  black-body curve of 500\,K. Some uncertainties in flux measurements are smaller than the used plot-symbol. The
  errorbars in wavelength indicate the width of the filter
  corresponding to 50\% transmission.}
\end{figure}

\noindent1.\,The dust emission is well approximated by a single 500\,K
Planck function,  strongly constrained by the WISE 22\,$\mu$m measurement.  The temperature of the warm dust is far lower than the dust sublimation
temperature of approximately 1500\,K. This implies that (1) orbits closer to the star are
(practically) devoid of dust grains otherwise dust excess emission
would be detected in the 2\,$\mu$m wavelength region; and (2) that the
dust distribution is limited in radial extent (a ring rather).
Assuming heating by stellar irradiation and dust particles in
thermal equilibrium with an optically thick environment one can
estimate the radial distance of the dust by applying $T_d \approx
  T_{\rm eff}(r/r_*)^{-1/2}$. Adopting the stellar parameters $T_{\rm
    eff}$ = 5250\,K and $R_*$ = 1.5\,R$_\odot$ (Paper\,I)\nocite{2010A&A...524A...8G} delivers a characteristic distance for
  the dust of $0.7$\,AU.  Dropping the optically
thick assumption and assuming optically thin dust instead, this characteristic distance reduces to 0.4\,AU. In addition, any 
cold (100\,K) dust component cannot be excluded by the current set of measurements,
but the total emission at 60\,$\mu$m should be $\la 0.05$\,Jy.

\noindent2.\,The second unusual property is the high fractional contribution of the
excess to the total luminosity which amounts to 8\%; this excess
places the star among the non-accreting objects with the strongest IR
excesses\footnote{A contribution $>$16\% was recently reported for the
  60\,Myr solar-type star V488 Per \citep{2012ApJ...752...58Z}.}.  We
can estimate the expected fractional contribution by the production of
small dust particles in a model for a steady-state, collisional
cascade of an asteroid belt, thought to be valid for debris disks
\citep{2007ApJ...658..569W}.  In brief, a steady state model adopts a
planetesimal size distribution that does not evolve, except that the
largest-sized bodies disappear by collisions and the minimum size is
set by stellar radiation pressure effects.  The cascade model
predicts a maximum infrared belt luminosity with time (eq. 18 in
  Wyatt 2008).  If the evolution of the planetesimal belt of RZ\,Psc
  is similar to that observed in A-type star debris disks (for which a
  fitting steady-state representation is parametrized by a maximum
  asteroid size of 60\,km, planetesimal strength of 150\,J/kg and
  eccentricity of 0.05) then with a characteristic distance of 0.7\,AU
  a belt width of 0.3\,AU, a fractional luminosity contribution of a
  few times $10^{-5}$\% is predicted at the adopted age of RZ\,Psc. This is a
  common value for debris disks. Much more mass in small dust grains
needs to be produced in RZ\,Psc than can be accounted for by the
steady-state collisional paradigm. We can use equation (4) from
Wyatt (2008) to convert the 8\% fractional contribution to a lower limit
on the mass, {\it viz.} $M_{\rm disk}> 1\cdot10^{23}$\,g. This conversion assumes
that the fractional luminosity defines the effective cross-sectional area of
the dust, i.e. an optically thin environment  (dust edge at 0.4\,AU). Additionally, we adopt
a single size and density for the dust particles, $10\,\mu$m and $\rm 3.3\,g\,cm^{-3}$
respectively (following Lawler \& Gladman 2012). However, if the same
dust is responsible for the excess emission {\it and} the UXOR variability, then
clearly the size of the dust should be (sub-)micron sized.

A dust belt with a sharp inner edge can be produced by dynamical
interactions between the dust and a secondary object located within
the gap between star and belt \citep{2011ARA&A..49...67W,2012ApJ...745....5K}.  If we assume this to be the case in RZ\,Psc, then 
we can derive the following about the secondary's properties. For a zero eccentricity orbit, 
the semi-major axis of the secondary has to be approximately half that
of the inner edge of the belt, (semi-major axis between 0.2 and 0.4\,AU) \citep{1994ApJ...421..651A}. The absence 
of any radial velocity variation  down to
$\rm 2\,km\,s^{-1}$ \citep{1993AstL...19..125S}, results in a secondary mass of $\la 38\,M_{\rm Jup}$ for 
optically thin dust (or $\la 53\,M_{\rm Jup}$ for the optically thick case). For a non-zero eccentricity 
orbit of a secondary object sculpting the inner rim of the dust distribution, the semi-major axis could be even 
smaller (e.g. Beust 2003\nocite{2003A&A...400.1129B}) and the mass upper limits are then also correspondingly lower. 

Alternative processes for explaining an inner-gap by
means of grain growth (as the first step in the planet formation
process) or photo-evaporation (responsible for the dissipation of the
primordial accretion disk, especially the disk's outer-parts) are
unlikely because they play a role in optically thick primordial disks
during the first 10\,Myr of the evolution \citep{2008ASPC..393...35C}.
\begin{figure}
\centering
\includegraphics[width=8.cm,height=9.5cm]{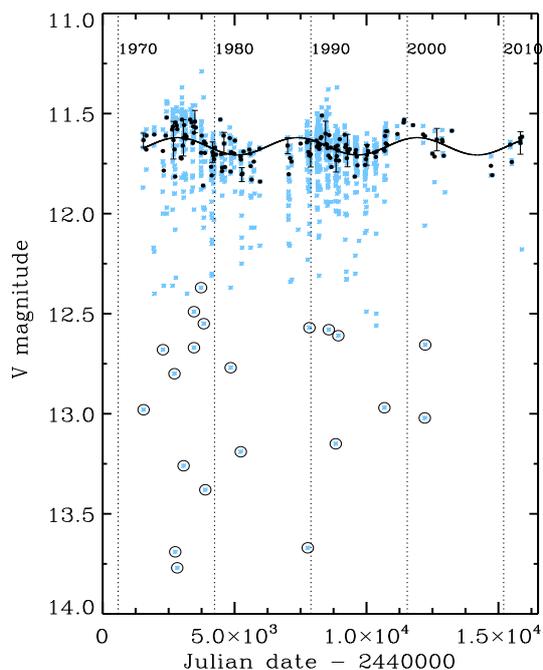}

\caption{The light curve of RZ\,Psc stretching 40 years. Over one
  thousand measurements are plotted (small blue stars). Gaps in the
  measurement coverage are caused by seasonal observability of the
  object on the sky. The deepest minima ($\rm \Delta V > 1$)
  are marked with open circles. Additionally, the flux at bright
  phases varies periodically. This periodicity is revealed by
  sigma-clipping and time-binning the data resulting in the points
  represented by black circles whose 12.44 year period is marked
by the sinusoid (see text). Typical uncertainties in the clipped-binned data
are indicated only in 10\% of the plotted points.}
\end{figure}

%\begin{figure}
%\begin{centering}
%\includegraphics[width=9cm, angle =0]{2.eps}
%\caption{\label{2} Schematic model of the nearest dust
%surroundings of RZ Psc (see text for more details). }
%\end{centering}
%\end{figure}

\subsection{The optical variability}
A large mass in small dust particles provides a reservoir for the
dust occultation events depicted in Fig.\,2.  The figure shows the
photometric activity of the star over the past forty
years. The light curve is based on our own observations and the
data from literature (Zaitseva 1985\nocite{1985PZ.....22..181Z};
Kardopolov, Sakhanenok \& Shutemova
1980\nocite{1980PZ.....21..310K}; Pugach
1981\nocite{1981Ap.....17...47P}; Kiselev et al. (1991);
Shakhovskoi et al. (2003). Two phenomena can be
identified:\newline (1) Brightness decreases of up to 2.5 visual
magnitudes. They occur on average once every year but
the events are aperiodic. The brightness minima last 1 to 2 days.
From the rate of flux change, one can estimate the
tangential velocity and approximate the distance for an opaque
screen (Paper\,I). From this, an orbital distance of
0.6\,AU is found, not inconsistent with the distance estimated from the
excess emission. The flux minima are accompanied by an increased
degree of polarization
\citep{1991Ap.....34..175K,2003ARep...47..580S}, which, taken
together, unambiguously identify small, micron-sized, dust grains
as the cause \citep{1991Ap&SS.186..283G}.  The occultation time
then limits the characteristic size for the clumps to $\rm \sim
0.05\,AU$.  The extinction at the beginning of the eclipses
displayed by RZ\,Psc is typical for UX Ori stars, and these can be
described approximately by assuming MRN size distribution for
sub-micron sized particles (Voshchinnikov et al. 1995\nocite{1995A&A...294..547V}).
Adopting the opacity at optical wavelengths for an ISM dust
mixture \citep{2000A&A...364..633N}, and with the observed
$\tau_{v}=3$ of the dust clump  (not an uncommon value for UXORs), we derive a mass of a few times
$10^{20}$\,g (assuming the clump consists of dust only).
\newline
(2) A modulation of the peak flux with a cycle of 12.4 years. The
brightness variability of RZ\,Psc has been presented in previous
publications reporting a quasi-periodic variation of the stellar
flux level during bright periods albeit with different periods
\citep{2003ARep...47..580S,1999AstL...25..243R}.  Here we present
a rigorous analytical period-search with the aim to settle the
question of any periodicity in RZ\,Psc's photometric data of the
past forty years.  

The light curve data present two challenges for
period searches. First, the data sampling is unevenly spaced in
time both due to seasonal observability and weather conditions.
Second, the brightness variability is dominated by an
intrinsically irregular component dominating the variability on a
time-scale of days, i.e. the obscuration of the star by the dust
clumps. The amplitude of the irregular variation is larger than
the measurement precision of less than approximately 0.03
magnitudes. We performed a period search on a subset of the data,
which was defined by discarding measurements deviating more than 2
$\sigma$ from the median brightness of the source. This step aims
to remove the signal caused by the obscuration events that lead to
the deep brightness minima. In the next step, the data were
averaged per time-bin with the aim to average out the small
amplitude of $\sim0.3^{m}$ daily variability of the source.
An optimal bin-size of 29 days was objectively calculated
according to a method which minimizes the bin-size by taking into
account the bin statistics (mean and variance) of the finite
sample \citep{shima}. The bin-averaged data points are represented
in Fig.\,2 by blue symbols and the errorbars  represent the
uncertainties taken to be the standard deviation per bin. If a
time-bin contains only one measurement then the average time-bin
standard deviation is assigned. Finally, a period search was
performed by computing the Generalized Lomb-Scargle (GLS)
periodogram \citep{2009A&A...496..577Z}.  This methodology
revealed a significant power at a period of 12.34 years with a
false alarm probability of $6.8\cdot10^{-4}$. No other significant
power peaks are present in the GLS periodogram. A subsequent
sinusoidal fit delivers a refined period of 12.44 years an
amplitude of 0.5 magnitude which is presented in Fig.\,2. Removal of
this periodic signal from the data results in a power-spectrum
without significant power beyond the noise.

The cyclic variability of cool stars can be connected with the
magnetic surface activity (magnetic cycles, see, e.g. Grankin et
al. 2008\nocite{2008A&A...479..827G}).  This type of activity is
frequently accompanied with the rotational brightness modulation of
stars not viewed pole-on (e.g. Vrba et
al. 1988\nocite{1988AJ.....96.1032V}). RZ Psc is observed close to
equator-on and its brightness does not display any rotation
modulation.  The cause for the long-term cyclic variability of RZ Psc
could be because of a warped disk (Grinin et
al. 2010\nocite{2010AstL...36..808G}) or large-scale perturbations in
a disk, as is suggested for UXORs (Grinin et
al. 1998\nocite{1998AstL...24..802G}).  Such perturbations can be
caused by the orbital motion of a co-planar low-mass companion
(Demidova et al. 2010\nocite{2010AstL...36..498D}). We speculate here,
that if this is so then the system is required to have a second
low-mass companion. Taking into account a 30\% uncertainty in star mass
and 0.25 year on the period, the semi-major axis of this 
tertiary component is $a=5.3\pm0.6$\,AU. We not that the application of this
particular model to RZ Psc is possible only if the outer part of the
disk (beyond the orbit of the second companion) contains some amount
of gas and fine dust. This material could be the remnant of the
primordial disk. This scenario would predict the existence of a far
infrared excess with $T \leq 100$\,K.  We underline that for a full
understanding of the RZ\,Psc system in particular the effects on the
dust distribution by the two inferred companion objects, numerical
modeling is required.

\begin{figure}
\includegraphics[width=8cm,angle=0]{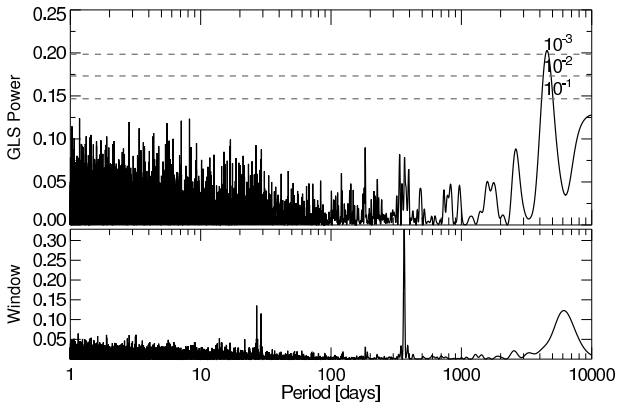}
\caption{Generalized Lomb-Scargle periodogram and window function. The computed
periodogram (upper plot) is a convolution of the astrophysical signal present in the data and
the sampling function present in these data (window function, lower plot). The periodogram is
obtained from the sigma-clipped, bin-averaged dataset. The horizontal dashed lines indicate
FAP levels of 10-1, 10-2, and 10-3. There is one significant peak present at a period of 12.34
years with a FAP of $6.8\cdot10^{-4}$. The window function shows a significant peak at 365 days
which is caused by the seasonal visibility of the source. However, significant alias peaks
caused by the sampling period are not present in the GLS periodogram. }
\end{figure}
\subsection{Warm debris disks}
RZ\,Psc adds to a growing number of main-sequence stars with
exceptionally strong, warm (500\,K) infrared excess. The excesses
are orders of magnitude stronger than can be explained by the
collisional cascade model applicable to regular debris disks
\citep{2005Natur.436..363S,2010ApJ...717L..57M,2012ApJ...752...58Z,2012ApJ...752...53L}.
Evidence is mounting that the warm debris disks are generated by
transient, or stochastic events based on the observed altered
emission properties of the dust grains
\citep{2012A&A...542A..90O}.  Proposed transient events constitute
a recent collision between two major bodies (like rocky planetary
embryos or even planets, Melis et al.
2010\nocite{2010ApJ...717L..57M}) or a second planetesimal belt at
larger radii feeding the inner dust distribution with evaporating
comets and/or inducing collisions, partially analogous to the
Solar system Kuiper belt \citep{2012A&A...542A..90O}. Involved
time-scales would favour the former explanation supported by the
fact that warm dust phenomenon near solar-type stars may occur
only in the first 30 to 100 Myr as judged from a sample of four
field stars \citep{2010ApJ...717L..57M}. Also young stellar
clusters indicate a maximum in the number stars that show
24\,$\mu$m excess emission around approximately 40\,Myr
\citep{2011MNRAS.411.2186S}. RZ\,Psc is unique in the sense that
the system's near equator-on orientation and dust occultations
give important clues to the dynamics of the system.

\section{Conclusions}
We have detected a strong mid-IR excess in the young star RZ\,Psc
with an 8\% fractional contribution to the bolometric luminosity
of the system.  The excess traces warm dust and it is well
described by a single Planck function of $\sim$500\,K. It
suggests a radially compact dust distribution  (a ring), with the inner-edge 
at 0.7 AU  if the dust is optically thick. We also 
find a periodicity of 12.4 years in the maximum optical brightness of the object. Although
uncertain, the estimates for the age indicate that the star is
beyond a formative phase and the mid-IR excess is unlikely to be
caused by a primordial disk. Therefore, copious amounts of small dust need to be
continually produced in this system to explain the optical
occultation events  (UXOR phenomenon), providing a dynamical view on the dust
production process.  The RZ\,Psc system 
could play a key-role in understanding the transition from 
primordial disks to debris disks.

\begin{acknowledgements} This publication makes use of data products from the Wide-field
   Infrared Survey Explorer, which is a joint project of the
   University of California, Los Angeles, and the Jet Propulsion
   Laboratory/California Institute of Technology, funded by the
   National Aeronautics and Space Administration and it makes use
   of data products from the Two Micron All Sky Survey, which is a
   joint project of the University of Massachusetts and the Infrared Processing and
   Analysis Center/California Institute of Technology, funded by the National
   Aeronautics and Space Administration and the National Science Foundation.
   This research is based on observations with AKARI, a JAXA project with the participation of ESA and
   has made use of NASA's Astrophysics Data System Bibliographic Services.
V.G. and I.P. were supported by grant of
the Presidium of RAS P21 and grant N.Sh.-1625.2012.2.
\end{acknowledgements}

%\bibliographystyle{aa}
%\bibliography{rz}

\end{document}